\documentclass[twocolumn,showpacs,preprintnumbers,amsmath,amssymb]{revtex4}
\usepackage{graphicx}% Include figure files
\usepackage{dcolumn}% Align table columns on decimal point
\usepackage{bm}% bold math

\def\Tr{\mathop{\rm Tr}\nolimits}

\begin{document}
%\draft

\title{Finite temperature lattice properties of graphene beyond the quasiharmonic approximation}

\author{K.V. Zakharchenko, M.I. Katsnelson, A. Fasolino}

\address {
Institute for Molecules and Materials, Radboud University
Nijmegen, Heyendaalseweg 135, 6525 AJ Nijmegen, The Netherlands
         }

\date{\today}

\begin{abstract}
The thermal and mechanical stability of graphene is important for many potential applications in nanotechnology.  We calculate the temperature  dependence  of lattice parameter, elastic properties and heat capacity by means of atomistic Monte Carlo simulations that allow to go beyond the quasiharmonic approximation. We predict an unusual, non-monotonic, behavior of the lattice parameter with minimum at $T \approx 900$ K and of the shear modulus with maximum at the same temperature. The Poisson ratio in graphene is found to be small $\approx 0.1$ in a broad temperature interval. 
\end{abstract}

\maketitle

Understanding the structural and thermal properties of two dimensional (2D) systems is one of the challenging problems in modern statistical physics~\cite{nelsonbook}. Traditionally, it was discussed mainly in the context of biological membranes and soft condensed matter. The complexity of these systems hindered any truly microscopic approach based on a realistic description of interatomic interactions. The discovery of graphene~\cite{kostya1}, the first truly 2D crystal made of just one layer of carbon atoms, provides a model system for which an atomistic description becomes possible. The interest for graphene has been triggered by its exceptional electronic properties (for review see~\cite{Geim1,Kats1,revCastro}) but the experimental observation of ripples in freely hanged graphene~\cite{Meyer} has initiated a theoretical interest also in the structural properties of this material~\cite{our,Castroneto}. Ripples or bending fluctuations have been proposed as one of the dominant scattering mechanisms that determine the electron mobility in graphene~\cite{KatsGeimPhilTrans}. Moreover, the structural state influences the mechanical properties that  are important in themselves for numerous potential applications of graphene~\cite{KatsGeimNanolet, Makianmicromechreson, Science}.

Two dimensional crystals are expected to be strongly anharmonic due to an intrinsic bending instability coupled to in-plane stretching modes. This coupling is crucial to prevent crumpling of the crystal and stabilize the flat phase~\cite{nelsonbook}. These expectations have been confirmed by atomistic simulations for graphene showing very strong bond length fluctuations already at room temperature~\cite{our}. Beside the relevance for 2D systems, anharmonicity~\cite{cowley} is of general importance in condensed matter in relation to structural phase transitions~\cite{trans1, trans2}, soft modes in ferroelectrics~\cite{soft}, melting~\cite{brat} and related phenomena. Usually anharmonicity in crystals is weak enough and thus can be well described in the framework of perturbation theory~\cite{cowley, vaks, zoli, kt}. However, this might be not the case for strongly anharmonic systems, like graphene. Atomistic simulations offer the possibility to study  anharmonic effects for a specific material without need of perturbative schemes. For carbon a very accurate description of energetic and thermodynamic properties of different allotropes including graphene~\cite{los2, our} is provided  by the empirical bond order potential LCBOPII~\cite{los1}. Here we present the temperature dependence of thermodynamical and elastic properties of graphene, calculated by means of atomistic Monte Carlo (MC) simulations based on LCBOPII. 

We perform MC simulations at finite temperature $T$ with periodic boundary conditions for a sample of $N=8640$ atoms with equilibrium size at zero temperature of 147.57~\AA~in the $x$ direction and 153.36~\AA~in the $y$ direction. We equilibrate the sample in the $NPT$ ensemble at pressure $P=0$ for at least $2\cdot10^5$ MC steps (1 MC step corresponds to $N$ attempts to a coordinate change) which we found to be enough for convergence of total energy and sample size. Further $10^5$ MC steps are used to evaluate the average lattice parameter $a$ and average nearest neighbor distance $R_{nn}$ and radial distribution function~$g(R)$.

\begin{figure}
\includegraphics[clip=true,width=0.9\linewidth]{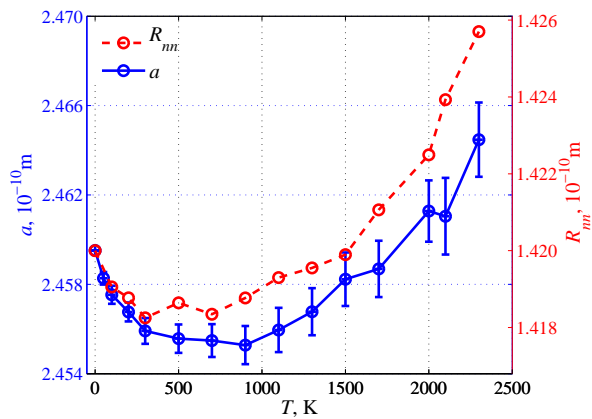}
\caption{(color online) Temperature dependence of the lattice parameter $a$ (solid blue line) and nearest neighbor distance $R_{nn}$ (dashed red line). The scales of left ($a$) and right ($R_{nn}$) $y$-axes are related to each other by $\sqrt{3}$. At $T=0$, $a=2.4595\cdot10^{-10}$~m.}
\label{fig1}
\end{figure}

\begin{figure}
\includegraphics[width=0.9\linewidth]{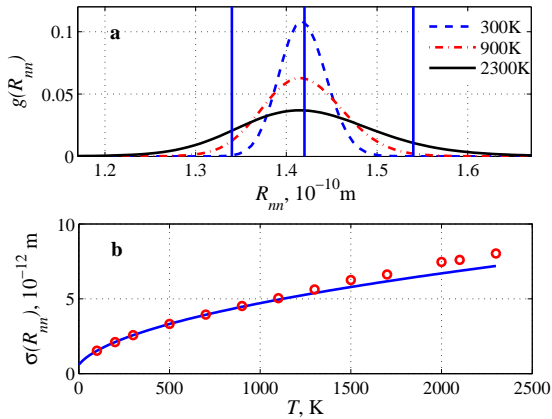}
\caption{a) (color online) Nearest neighbor radial distribution function $g(R_{nn})$ for the $N = 8640$ sample at 300~K, 900~K and 2300~K. The vertical lines indicate the length of double ($1.34\cdot10^{-10}$~m), conjugated ($1.42\cdot10^{-10}$~m) and single ($1.54\cdot10^{-10}$~m) bonds. b) Standard deviation $\sigma (R_{nn})$ (red circles) and the best fit to $\sqrt{T}$ in the temperature range up to 500~K (solid blue line).}
\label{fig2}
\end{figure}

Figure~\ref{fig1} shows that $a$  and $R_{nn}$ decrease with increasing temperature up to about 900~K, yielding a negative thermal expansion coefficient $\alpha = (-4.8\pm1.0)\cdot10^{-6}~{\rm K}^{-1}$ in the range 0--300~K. As noted in Ref.~\cite{MoMar} this anomaly is due to a low-lying bending phonon branch~\cite{Lifshitz}. Our results are in agreement up to 500~K with those of Mounet and Marzari~\cite{MoMar} who used  the quasi-harmonic approximation with phonon frequencies and Gr\"uneisen parameters calculated by the density functional approach.  However, at higher temperatures our results are qualitatively different, since in Ref.~\cite{MoMar} $\alpha$\ remains negative in the whole studied temperature interval up to 2200~K, whereas we find that it changes sign and becomes positive at $T\approx 900$ K. This discrepancy with the quasi-harmonic theory, which in general works reasonably well for three-dimensional crystals, is one of the evidences of strong anharmonicity in graphene.

The deviations from harmonic behavior can be characterized by examining the radial distribution function $g(R)$ around the first neighbor distance $R_{nn} = 1.42\cdot10^{-10}$~m. In Fig.~\ref{fig2}(a) we present $g(R)$ and the related standard deviation $\sigma (R_{nn})$ shown in Fig.~\ref{fig2}(b). In the harmonic approximation $R_{nn}$ would have a Gaussian distribution yilding $\sigma (R_{nn}) \propto \sqrt{T}$. Deviations from square root behavior can be observed above 900~K, achieving 10~\% at 2000~K.

The Lindemann criterion has been shown to apply also in 2D, giving $\sigma (R_{nn}) \approx 0.23 R_{nn}$ at melting~\cite{Lindemann}. We found $\sigma(R_{nn})/R_{nn}=0.056$ at $T=2300$~K, indicating that we are significantly below melting point. Moreover, conventional theory of two-dimensional melting relates it to the formation of topological defects~\cite{Halperin}. In our simulations we have not seen any sign of premelting anomalies (formation of vacancies, topological defects etc.) up to 3500~K~\cite{our}.

\begin{figure}
\includegraphics[clip=true,width=0.9\linewidth]{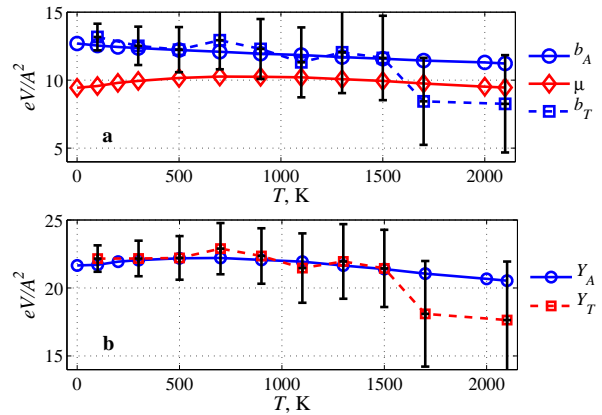}
\caption{a) (color online) 2D elastic moduli of graphene as of function of temperature: adiabatic bulk modulus $b_A$ (solid blue line with circles), isothermal bulk modulus $b_T$ (dashed blue line with squares) and shear modulus $\mu$ (solid red line with diamonds). b) Adiabatic Young's modulus $Y_A$ (solid blue line with circles) and isothermal $Y_T$ (dashed red line with squares).}
\label{fig5}
\end{figure}

\begin{figure}
\includegraphics[clip=true,width=0.9\linewidth]{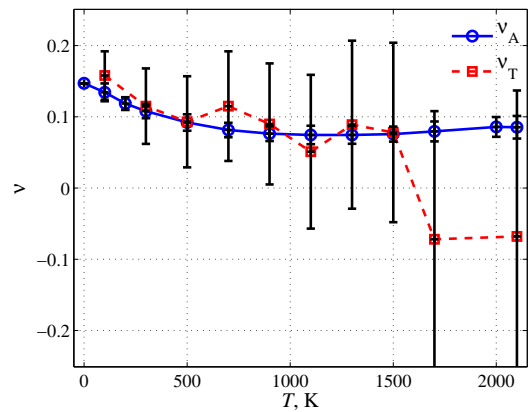}
\caption{Adiabatic $\nu_A$ and isothermal $\nu_T$ Poisson ratio of graphene as function of temperature.}
\label{fig7}
\end{figure}

The strong anharmonic behavior of graphene leads also to unusual temperature dependence of the elastic moduli.

The 2D bulk modulus $b$ is defined by
\begin{equation} \label{eq:F_uniform1}
E_{is} = 2 b u_{is}^2,
\end{equation}
where $E_{is}$ is the elastic energy per  unit area under an isotropic deformation  $u_{yy} = u_{xx}= u_{is},\ u_{xy}=0$. 

For uniaxial deformations $u_{xx}$ ($u_{yy}=u_{xy}=0$) the elastic energy is
\begin{equation} \label{eq:F_unidir1}
E_{uni} = \frac{1}{2} (b+\mu) u_{xx}^2,
\end{equation}
where $\mu$ is the 2D shear modulus. 

Isothermal moduli are also expressed as in Eq.~(\ref{eq:F_uniform1}) and Eq.~(\ref{eq:F_unidir1}), with replacement of the energy $E$ by the free energy $F=-T\ln{Z}$ where $Z$ is the partition function. Although it is impossible in MC to calculate $F$ directly, we will use the fact that  adiabatic and isothermal shear moduli $\mu$ coincide~\cite{Landau} and that the Poisson ratio defined below can be calculated directly to derive the isothermal bulk modulus $b_T$. The Young modulus $Y$ and Poisson ratio $\nu$ are defined in terms of $b$ and $\mu$~as~\cite{Chaikin}

\begin{align} \label{eq:Young_bulk_shear}
Y &= \frac{4 b \mu}{b+\mu},\\[1ex]
 \label{eq:Poisson_eq}
\nu &= \frac{b - \mu}{b+\mu}.
\end{align}

The Poisson ratio can also be defined as the ratio between the axial $\epsilon_{axial}$ and transverse $\epsilon_{trans}$ strain as

\begin{equation} 
 \label{eq:Poisson_eq2}
\nu = -\frac{\epsilon_{trans}}{\epsilon_{axial}}.
\end{equation}

The latter definition provides a way to calculate the isothermal $\nu_T$ so that Eq.~(\ref{eq:Poisson_eq}) with $\mu_A=\mu_T=\mu$ yields~$b_T$.

Adiabatic bulk and shear moduli $b_A$ and $\mu$ have been calculated using the following procedure. We equilibrate the sample as described before.  Afterwords, 20 configurations separated by 5000~MC steps were stored and subjected to either isotropic or uniaxial deformation in steps of 0.01~\% without letting the sample relax. For each sample, the variation of the elastic energy with deformation was then fitted to Eq.~(\ref{eq:F_uniform1}) and Eq.~(\ref{eq:F_unidir1}) over 21 points around the undistorted configuration. The averages of the calculated  $b_A$ and $\mu$ for the 20 samples are given in Table I and shown in Fig.~\ref{fig5} together with the derived $Y_A$. We find that the temperature dependence of $\mu$ is anomalous. While in general all elastic moduli decrease as a function of temperature due to weakening of interatomic interactions with temperature, in graphene $\mu$ grows with increasing temperature up to $T\sim 700-900$~K which is the same temperature where the thermal expansion behavior (Fig.~\ref{fig1}) becomes normal. The Young modulus $Y$ follows the same anomalous temperature dependence as~$\mu$.

We find that the  behavior of the elastic energy as a function of deformation $u$  is parabolic  in a wider range of deformations, up to about 0.2~\%. For larger deformations, the elastic energies follow a cubic dependence on the deformation at least up to $u=3~\%$. At this value the ratio of the cubic term to the quadratic one in the elastic energy is about $0.12$. Up to 10~\% deformation and up to 2200 K, deformations are reversible, and no defect (vacancy and Stone-Wales~\cite{Carlsson} or dislocations~\cite{Vozmediano}) are found. This is not surprising in view of the very high cohesive energy (7.6 eV/atom in graphite~\cite{los1}) of carbon and defect formation energy in graphene~\cite{Carlsson}. To the best of our knowledge, there are no experimental data on defect formation under strain in this range of temperatures. 

\begin{table}
\caption{\label{tab:table1}Adiabatic bulk ($b_A$), shear ($\mu$) and isothermal bulk ($b_T$) moduli, and isothermal Poisson ratio ($\nu_T$).}
\begin{ruledtabular}
\begin{tabular}{cccccc}
$T$, K & $b_A~\rm (eV\cdot\AA^{-2})$ & $\mu~\rm (eV\cdot\rm\AA^{-2}$) & $b_T~\rm (eV\cdot\rm\AA^{-2})$ & $\nu_T$\\
\hline
0& 12.69 & 9.44 & -- & --\\
100& 12.54$\pm$0.05 & 9.57$\pm$0.21 & 13.17$\pm$0.98 & 0.16$\pm$0.03 \\
200& 12.44$\pm$0.03 & 9.80$\pm$0.15 & -- & -- \\
300& 12.36$\pm$0.04 & 9.95$\pm$0.17 & 12.52$\pm$1.41 & 0.12$\pm$0.05 \\
500& 12.22$\pm$0.05 & 10.16$\pm$0.20 & 12.24$\pm$1.66 & 0.09$\pm$0.06 \\
700& 12.09$\pm$0.05 & 10.27$\pm$0.17 & 12.93$\pm$2.13 & 0.12$\pm$0.08 \\
900& 11.94$\pm$0.04 & 10.25$\pm$0.18 & 11.29$\pm$2.20 & 0.09$\pm$0.09 \\
1100& 11.85$\pm$0.06 & 10.21$\pm$0.22 & 11.31$\pm$2.57 & 0.05$\pm$0.11 \\
1300& 11.70$\pm$0.04 & 10.07$\pm$0.21 & 12.05$\pm$3.00 & 0.09$\pm$0.12 \\
1500& 11.57$\pm$0.04 & 9.94$\pm$0.18 & 11.63$\pm$3.10 & 0.08$\pm$0.13 \\
1700& 11.44$\pm$0.04 & 9.75$\pm$0.24 & 8.44$\pm$3.20 & -0.07$\pm$0.18 \\
2000& 11.31$\pm$0.06 & 9.52$\pm$0.22 & -- & -- \\
2100& 11.23$\pm$0.05 & 9.46$\pm$0.26 & 8.26$\pm$3.58 & -0.07$\pm$0.21 \\
\end{tabular} \end{ruledtabular} \end{table}

Next, the isothermal Poisson ratio $\nu_T$ has been calculated using the following procedure. We take the graphene sample equilibrated as described before at a given temperature. The sample is then stretched of 1~\% in the $x$ and $y$ directions separately and re-equilibrated again for at least $5 \cdot 10^4$~MC~steps. After re-equilibration, the sample size in the $x$ and $y$ directions have been averaged for at least $5\cdot10^4$~MC~steps and the corresponding strain $\epsilon_x$ and $\epsilon_y$  have been calculated yielding the Poisson ratio in each direction through Eq.~(\ref{eq:Poisson_eq2}). The Poisson ratios in the $x$ and $y$ directions are very close and we take their average as  $\nu_T$. The calculated adiabatic and isothermal Poisson ratios $\nu_A$ and $\nu_T$, shown in Fig.~\ref{fig7} and Table I, are very small and coincide within the error in the whole studied temperature range. However at high temperature, we find that $\nu_T$ can become negative. Materials with negative Poisson ratio are called  auxetic and, in general, this property is related to very unusual crystalline structures. Membranes, on the other hand, may display this behavior due to entropy. In fact, an expansion in the unstretched direction contrasts the reduction of phase space due to the decrease of height fluctuations due to stretching. Furthermore, the smallness of $\nu$ implies that the Lame' constant $\lambda=b-\mu$ is small in comparison with $\mu$. Therefore for a generic deformation described by a tensor $\hat{u}$, the elastic energy $E_{el}= \mu u_{ij}^2+(1/2) \lambda (\Tr \hat{u}^2)$ ~\cite{nelsonbook} for graphene can be approximated as $E_{el}\approx \mu u_{ij}^2$.

Once $\nu_T$ is known we can calculate $b_T$ from  Eq.~(\ref{eq:Poisson_eq}) and $Y_T$ from Eq.~(\ref{eq:Young_bulk_shear}). The calculated $b_T$ and $Y_T$ are presented in Table I and compared to the adiabatic values in Fig.~\ref{fig5}. At $T=300~\rm~$K, we find $Y_A=353 \pm 4 \rm~N \cdot m^{-1}$ and $Y_T=355\pm21 \rm~N \cdot m^{-1}$ in good agreement with the experimental value $340\pm50 \rm~N \cdot m^{-1}$~\cite{Science}.

\begin{figure}
\includegraphics[clip=true,width=0.9\linewidth]{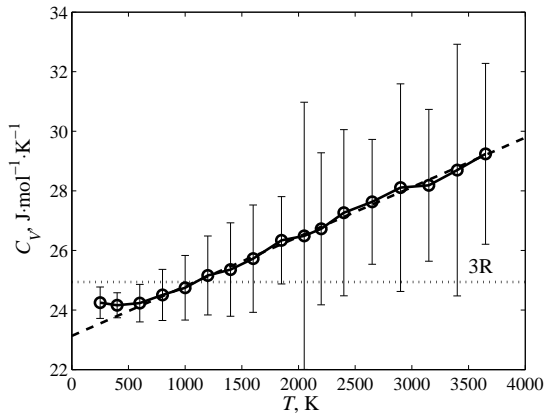}
\caption{Temperature dependence of the molar heat capacity at constant volume $C_V$ (solid line) and fit to $C_V(T)=3R (1+T/E_0)$ (dashed line).}
\label{fig8}
\end{figure}

Another important anharmonic effect is the temperature dependence of the molar heat capacity at constant volume $C_V(T)=3R (1+T/E_0)$ (see Fig.~\ref{fig8}), where $R$ is the gas constant and $E_0$ is a typical energy of interatomic interactions~\cite{cowley, kt}. The low temperature behavior was calculated in the harmonic approximation in~\cite{MoMar}. Our approach is classical and therefore can be used to calculate $C_V$ only at high temperatures. On the other hand, our approach does not use the harmonic approximation, yielding information about phonon-phonon interaction effects. Our calculations show that the linear temperature dependence of $C_V$ becomes noticeable for $T> 800$ K with $E_0=1.3$~eV. Contrary to alkali metals where $E_0$ is of the order of the vacancy formation energy~\cite{vaks}, for graphene, due to anharmonicity, $E_0$ is about $1/5$ of the defect formation energy.

In summary we have presented the temperature dependence of lattice parameter, elastic moduli and high temperature heat capacity of graphene calculated by Monte Carlo simulations based on the LCBOPII empirical potential \cite{los1} for a crystallite of about 15$\times$ 15 nm$^2$.  In the studied range of temperatures, up to $2200$~K, and for deformation as large as $10\%$ we have not seen any sign of defect formation. Indeed the very high energy for defect formation in graphene makes this material exceptionally strong, as also found experimentally~\cite{Science, KatsGeimNanolet}. We find that graphene is strongly anharmonic due to soft bending modes yielding strong out of plane fluctuations. We find  that, up to $900$~K, graphene is anomalous since its lattice parameter decreases and shear modulus increases with increasing temperature going over to normal behavior at higher temperatures. It would be interesting to check these predictions experimentally. 

We thank Jan Los for discussions. This work is part of the research programme of the 'Stichting voor Fundamenteel Onderzoek der Materie (FOM)', which is financially supported by the 'Nederlandse Organisatie voor Wetenschappelijk Onderzoek (NWO)'.

\end{document}